\documentclass[aps,prl,reprint]{revtex4-1} 

\usepackage{graphicx} 
\usepackage{amsmath,amssymb}
\usepackage{float}
\usepackage{units}

\newcommand{\D}{\displaystyle}

\begin{document} 

\title{Solution of the proton radius puzzle?\\
Low momentum transfer electron scattering data are not enough.} 
 
\newcommand{\kph}{\affiliation{Institut f\"ur Kernphysik, Johannes
Gutenberg-Universit\"at Mainz, 55128 Mainz, Germany.}}
\newcommand{\lns}{\affiliation{Laboratory for Nuclear Science, MIT,
Cambridge, Massachusetts 02139, USA.}}

\author{Michael O.\ Distler}\kph 
\author{Thomas Walcher}\kph
\author{Jan C.\ Bernauer}\lns

\date{November 2, 2015}  

\begin{abstract} 
In two recent papers it is argued that the ``proton radius puzzle''
can be explained by truncating the electron scattering data to low
momentum transfer and fit the rms radius in the low momentum expansion
of the form factor. It is shown that this procedure is inconsistent
and violates the Fourier theorem. The puzzle cannot be explained in
this way.
\end{abstract} 
 
 
\maketitle  

The ``proton radius puzzle'' is the difference of the rms radius $R_p
= \langle r^2 \rangle^{1/2}$ as determined from elastic electron
scattering and as derived from a very precise Lamb shift measurement
of muonic hydrogen. The electron scattering result $R^e_p =
\unit[0.877(5)]{fm}$
\cite{Bernauer:2011zz,Bernauer:2013tpr,Arrington:2015ria} deviates
from the muonic result $R^\mu_p = \unit[0.8409(4)]{fm}$ by
$\unit[0.036(5)]{fm}$ \cite{Pohl10,Antognini:1900ns} or 7 standard
deviations.

In two recent papers Keith Griffioen, Carl Carlson, and Sarah Maddox
\cite{Griffioen:2015hta} and Douglas W.\ Higinbotham, et al.\
\cite{Higinbotham:2015rja} conjecture that this difference, sometimes
called the ``proton radius puzzle'', could be resolved by just
restricting the analysis of the electron scattering data to the data
at low momentum transfer $Q^2$. In ref.\ \cite{Griffioen:2015hta} the
new data of Bernauer et al.\ for $Q^2 < \unit[0.02]{(GeV/c)^2}$
\cite{Bernauer:2011zz,Bernauer:2013tpr} are used, whereas in
ref.\ \cite{Higinbotham:2015rja} the old data, i.e.\ before 1980, for
$Q^2 < \unit[0.03]{(GeV/c)^2}$ \cite{Murphy:1974zz,Simon:1980hu} are
analyzed. In both analyses radii are extracted which appear close to
the one derived from the muonic Lamb shift within relatively large
errors. In both papers the statistical significance of these results
are in focus, however, it is overlooked that the approach of limiting
the data sets to small $Q^2$ is not in accord with basic facts of form
factors and rms radii derived from them as will be shown in the
following.

The charge distribution of a nucleus and its form factor are connected
by the 3-dimensional Fourier transforms
\begin{equation}
\rho(r) = \frac{1}{(2 \pi \hbar \,c)^3} \int d^3Q \, G(Q^2) \,
\exp(\frac{i}{\hbar \,c} \, \vec{Q} \cdot \vec{r} )
\label{eq:rho}
\end{equation}
and its inversion
\begin{equation}
G(Q^2) = \int d^3r \, \rho(r) \, \exp(- \frac{i}{\hbar \,c} \, \,
\vec{Q} \cdot \vec{r})
\label{eq:G}
\end{equation}
This implies that either $G(Q^2)$ or $\rho(r)$ is known and integrated
over the full range of $0 \leqslant Q^2 \leqslant \infty$ or $0
\leqslant r \leqslant \infty$, respectively. Experimentally the
minimum and maximum of $Q^2$ are limited and $G(Q^2)$ cannot be
determined over the full range. Therefore, one has to use models for
$\rho(r)$ for nuclei, or for $G(Q^2)$ for the proton. However, it is
mandatory to ensure that the models used are consistent with the
Fourier theorem given by eqs.\ (\ref{eq:rho}) and (\ref{eq:G}).

The method used in refs.\ \cite{Griffioen:2015hta,Higinbotham:2015rja}
is well known since the early days of electron scattering and uses the
expansion of eq.\ (\ref{eq:rho}):
\begin{equation}
G(Q^2) = 1- \frac{1}{6} \langle r^2 \rangle Q^2 + \frac{1}{120}
\langle r^4 \rangle Q^4 - \frac{1}{5040} \langle r^6 \rangle Q^6
+-\cdots
\label{eq:exp}
\end{equation}
where
\begin{equation}
 \langle r^n \rangle = 4\pi \int r^2 dr \, \rho(r) \, r^n .
 \label{eq:rms}
 \end{equation}
In the two papers the rms radius in eq.\ (\ref{eq:rms}) is determined
by fitting the truncated data basis for low $Q^2$ by
eq.\ (\ref{eq:exp}) to order $Q^2$ (linear), $Q^4$ (quadratic), and
$Q^6$ (cubic) \cite{Griffioen:2015hta} and order $Q^2$
\cite{Higinbotham:2015rja}. It is a well known fact though, that a
sharp truncation in the coordinate space (e.g. a uniformly charged
sphere) produces an oscillating behaviour in momentum space. The same
is true for a sharp truncation in momentum space.  A ``saw tooth''
like form factor as given by eq.\ (\ref{eq:sawG}) corresponds to the
charge distribution as given by eq.\ (\ref{eq:sawrho}), i.e.\ produces
an oscillating behaviour in coordinate space.

\begin{equation}
G(q)=\left(1-\frac{q^2 R^2}{6 (\hbar c)^2}\right) \Theta
  \left(\frac{6 (\hbar c)^2}{R^2}-q^2\right)
  \label{eq:sawG}
  \end{equation}
  
\begin{figure}[h]
\centering
\includegraphics[width=0.85\columnwidth]{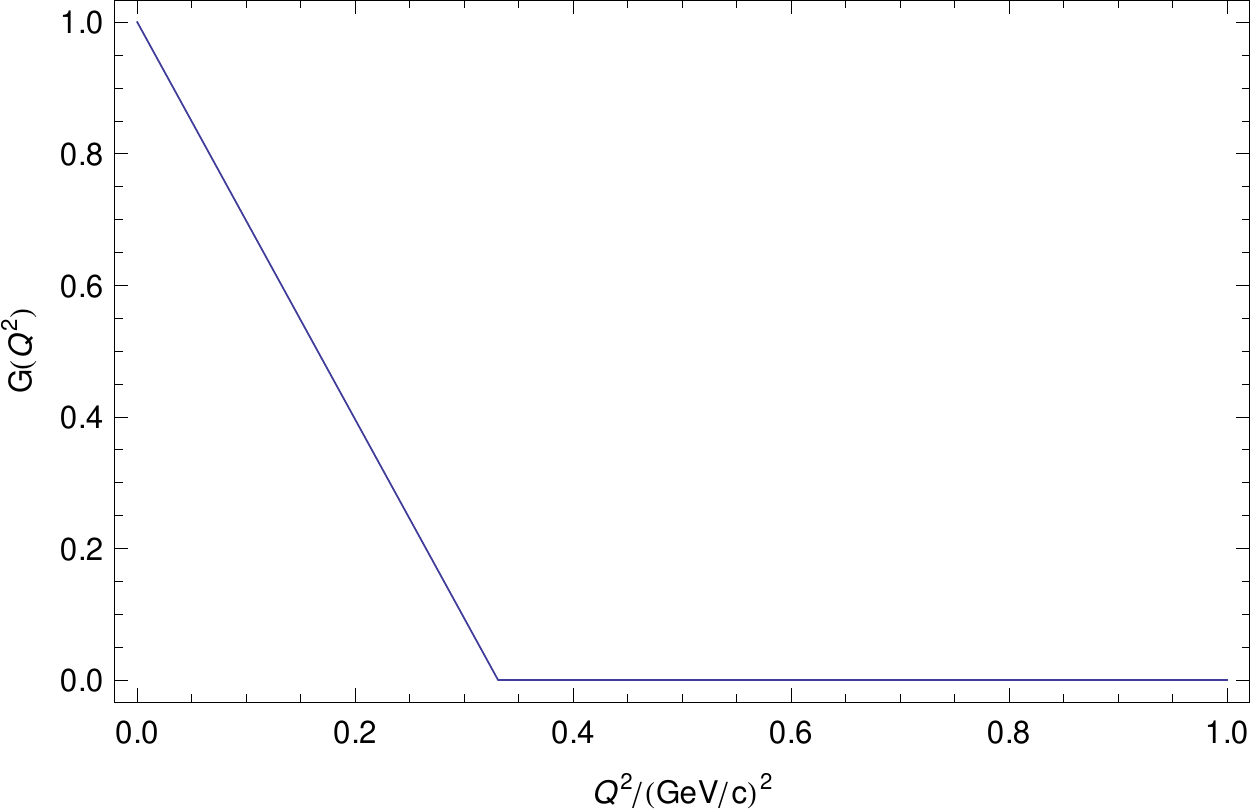}
\caption{The form factor $G(Q^2)$ for the ``saw tooth'' model. }
\label{fig:sawG}
\end{figure}

\begin{equation}
\rho(r)=-\frac{\left(2 r^2-R^2\right) \sin \left(\frac{\D \sqrt{6}\,
  r}{\D R}\right)+\sqrt{6} \,r R \cos \left(\frac{\D \sqrt{6}\,
  r}{\D R}\right)}{2 \pi ^2 r^5}
\label{eq:sawrho} 
\end{equation} 

\begin{figure}[h]
\centering
\includegraphics[width=0.85\columnwidth]{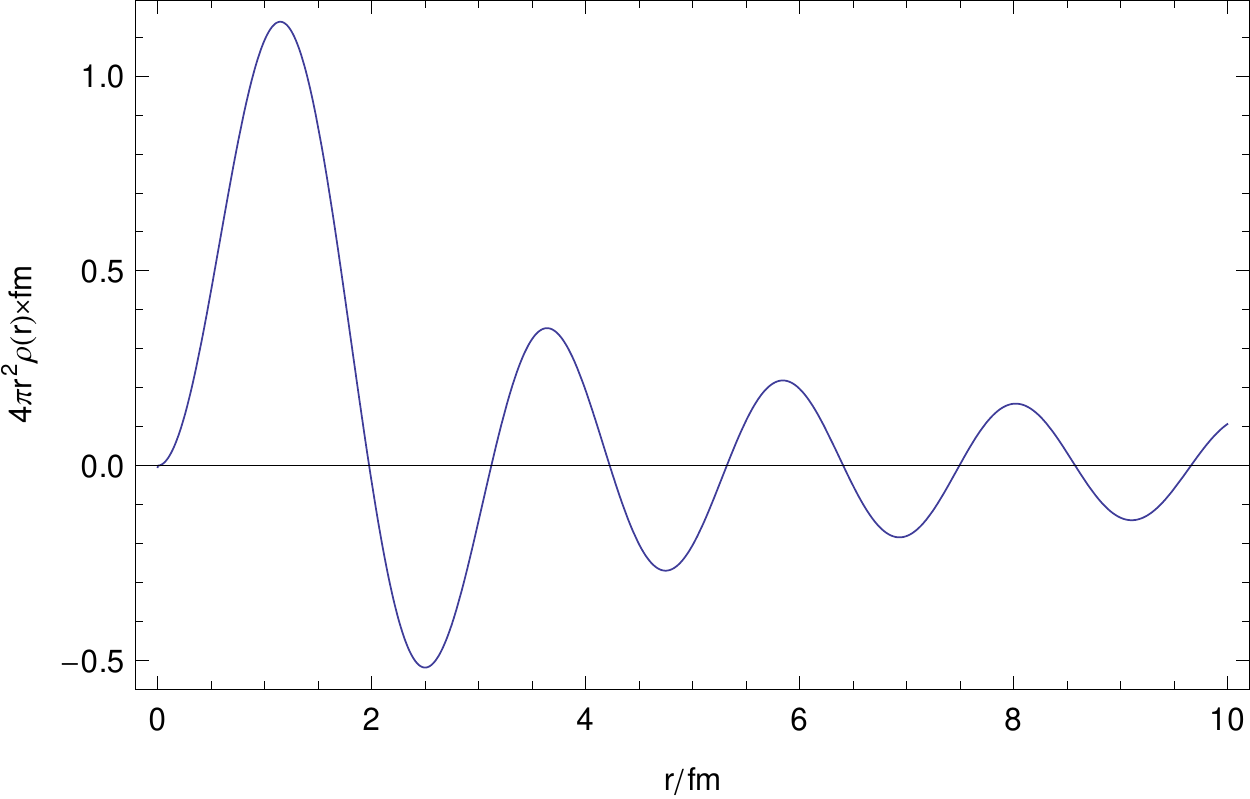}
\caption{The charge distribution $\rho(r)$ for the ``saw tooth'' model. }
\label{fig:sawrho}
\end{figure}

\begin{table}
\setlength{\arraycolsep}{1pt}
$$\begin{array}{l|ccccc}
\rho(r) \hbox{ label}
   & \D \frac{\langle r^2 \rangle}{\hbox{fm}^2}
   & \D \frac{\langle r^4 \rangle}{\hbox{fm}^4}
   & \D \frac{\langle r^6 \rangle}{\hbox{fm}^6}
   & \D \frac{\langle r^1 \rangle_{(2)}}{\hbox{fm}}
   & \D \frac{\langle r^3 \rangle_{(2)}}{\hbox{fm}^3} \nonumber \\[7pt] \hline
\hbox{Exponential~} & 0.7056 & 1.2447 & 4.0985 & 1.0609 & 2.2457 \nonumber \\
\hbox{Gaussian} & 0.7056 & 0.8298 & 1.3662 & 1.0945 & 2.0594 \nonumber \\
\hbox{Uniform} & 0.7056 & 0.5927 & 0.5421 & 1.1154 & 1.9433 \nonumber \\
\hbox{Saw tooth} & 0.7056 & 0 & 0 & 0.7277 & 0.6162 \nonumber \\
\hbox{Experiment} & 0.774 & 2.59 & 29.8 &
1.085 & 2.85  \nonumber \\
\hbox{(stat.)(syst.)\,err.} & (8) & (19)(04) & (7.6)(12.6) & (3) & (8) \nonumber 
\end{array}$$
\caption{Moments of the indicated charge distributions also including
the first and third Zemach moments. The moments are given for an rms
radius of $\unit[0.840]{fm}$. Experimental values are taken from
\cite{Distler:2010zq}. All charge distributions are equivalent to
simple form factor models which are evidently at variance with the
experimentally observed form factor of the proton. The saw tooth model
demonstrates clearly the impossibility to neglect the high $Q^2$
range.}
\label{tbl:expval}
\end{table}

\noindent
For this form factor all values $\langle r^n \rangle$ for $n\geqslant
4$ obtained with the derivatives of eq.\ (\ref{eq:exp}) are zero.  It
produces also unreasonably small values for the Zemach moments (see
Tab.\ \ref{tbl:expval}) inconsistent with both electron scattering and
spectroscopy \cite{Antognini:1900ns,Carlson:2015jba}. This is of
course not observed and unphysical as there is no known mechanism that
could produce oscillating charges at large distances in the proton.
However, this pathological situation is within the statistical errors
of the fits of Carl Carlson \cite{Carlson:2015jba} to the truncated
data range taking $\langle r^2 \rangle$ and $\langle r^4 \rangle$ as
fit parameters and with Higinbotham et al.\ \cite{Higinbotham:2015rja}
putting unjustifiably $\langle r^4 \rangle \equiv 0$.

In ref.\ \cite{Griffioen:2015hta} the problem is realized and its
influence is determined for three of the charge distribution in the
Tab.\ \ref{tbl:expval}: exponential, Gaussian and uniformly charged
sphere. But these are just very crude models. We know more about the
proton. Two independent studies of the world data have been published
by John Arrington et al.\ \cite{Arrington:2007ux,Arrington:2015ria}
and Jan Bernauer et al.\ \cite{Bernauer:2011zz,Bernauer:2013tpr} based
on the measured form factors up to large $Q^2 \leqslant
\unit[10]{(GeV/c)^2}$. It should be unnecessary to state that models like
the Gaussian or homogeneously charged sphere are excluded for the
proton since the work of Robert Hofstadter \cite{Hofstadter:1956abc}.

The extracted radius in ref.\ \cite{Griffioen:2015hta} depends
strongly on the $Q^4$ term, the curvature of $G(Q^2)$ at small $Q^2$:
higher values $\langle r^4 \rangle$ correlate with larger radii
$\langle r^2 \rangle$. The fits over the whole $Q^2$ range by Bernauer
et al.\ \cite{Bernauer:2011zz,Bernauer:2013tpr} indeed find a strong
curvature. In Kraus et al.\ \cite{Kraus:2014qua} it has been shown that a
low-order fit including a fit of the curvature to a truncated data
set is not reliable.

It is noted that the form factor and charge distributions going into
the rms radius in the two independent studies
\cite{Arrington:2007ux,Arrington:2015ria,Bernauer:2011zz,Bernauer:2013tpr}
are automatically fulfilling the Fourier relation, since they are
determined from one consistent fit over the full $Q^2$ range. The two
papers \cite{Griffioen:2015hta,Higinbotham:2015rja} using the
truncated $G(Q^2)$ disregard this consistency. Griffioen et al.\
\cite{Griffioen:2015hta} insert the rms radius fitted for small $Q^2$
into a model assumed to be valid for large $Q^2 \geqslant
\unit[0.02]{(GeV/c)^2}$. However, this form factor is not in agreement
with the cited measurements
\cite{Arrington:2007ux,Arrington:2015ria,Bernauer:2011zz,Bernauer:2013tpr}.

Higinbotham et al.\ \cite{Higinbotham:2015rja} neglect the $Q^4$ term
completely though they fit the data of the larger region $Q^2
\leqslant \unit[0.03]{(GeV/c)^2}$.  Yet, their reasoning is erroneous on
several aspects. First, we know that the {\it true} form factor has a
finite curvature, so using an F-test to decide about the significance
of the $Q^4$-term in the expansion of eq.\ (\ref{eq:exp}) is not
justified. Any hypothesis test in classical statistics is
based on a very important assumption: one has to know the {\it true}
function. It may be the case that we do not know the precise true
functions for the form factors of the proton, but the truncated
polynomial of eq.\ (\ref{eq:exp}) can definitely be excluded.  A
p-value or a significance level calculated with a wrong model
assumption is not valid.

Second, large off-diagonal elements in the covariance matrix are not a
``problem''. On the contrary, if one fits a polynomial expansion like
the one in eq.\ (\ref{eq:exp}) one will always end up with highly
correlated parameters.  It is also always possible to construct an
orthogonal functional basis (ref.\ \cite{James:2006st} recommends the
Forsythe method for polynomials) where the resulting covariance matrix
of the parameters is indeed diagonal. For a quantitative discussion of
the correlation it is useful to calculate the correlation matrix:
$$\rho_{ij}=\frac{\sigma_{ij}}{\sigma_i\sigma_j}
\qquad\hbox{with}\,-1\le\rho_{ij}\le 1$$
where the $\sigma_{ij}$ are the elements of the covariance matrix and
$\sigma_{i}=\sqrt{\sigma_{ii}}$. For the covariance matrix in
\cite{Higinbotham:2015rja}, eq.\ (5) therein, one gets
$\rho_{23}=-0.95$ and therefore a strong negative correlation between
the linear and the quadratic parameter. Taking into account the factor
$-1/6$ of eq.\ (\ref{eq:exp}) one gets a large positive correlation
between the quadratic parameter and the extracted radius. The authors
of \cite{Higinbotham:2015rja} have shown themselves that if they
artificially reduce the quadratic term to zero, they reduce the radius
from $R_p=\unit[0.875]{fm}$ to $\unit[0.840]{fm}$ therefore giving a
very strong reason not to neglect the quadratic term.

But the reasoning in ref.\ \cite{Higinbotham:2015rja} has a more
serious problem still. In their F-test two regression models are compared,
where one model, the order 2 polynomial, includes the second, linear,
model.  It is clear that the order 2 polynomial will always fit the
data better than the linear model, unless the quadratic term becomes
zero and both models are identical. So, Higinbotham et al.\ are
rejecting a model not because it gives a worse fit but because the fit
is not ``significantly'' better. Moreover, fitting the data well is
only a precursor to the more important goal: getting a robust estimate
of the rms radius. As we have argued in the previous paragraph, the
extracted radius changes dramatically when the order of the polynomial
model is reduced from quadratic to linear. This makes the quadratic
term very significant for the extraction of the radius.

In addition, their two parameter fits are not influenced by the form
factor at large $Q^2$. Therefore, their considerations of this form
factor in the second part of their paper are in view of refs.\
\cite{Arrington:2007ux,Bernauer:2011zz,Bernauer:2013tpr} not only
misplaced, but wrong.

A numerically precise calculation with the charge distribution derived
from eq.\ (\ref{eq:rho}) and based on the data and fits of refs.\
\cite{Arrington:2007ux,Bernauer:2011zz} of moments with $n \leqslant
6$ and Zemach moments is published in ref.\ \cite{Distler:2010zq}.

One may ask why the method of small $Q^2$ expansion was relatively
successful for nuclei. This is due to the fact that the short range
nuclear force produces to a good approximation a uniformly charged
sphere and one can derive the $R_A \propto A^{1/3}$ dependence of the
rms radius. This is a good model for nuclei. However, for the charge
distribution of the proton we do not yet have a good model and
consequently the form factor and the rms radius are only derivable
from a fit over a sufficiently wide $Q^2$ range as performed in refs.\
\cite{Arrington:2007ux,Arrington:2015ria,Bernauer:2011zz,Bernauer:2013tpr}.
Ingo Sick has investigated the minimal $Q^2$ required in a recent
paper \cite{Sick:2014kna}.

The approach of the two papers is, nonetheless, not only wrong in
analytical terms it also misunderstands the statistical evaluation of
physics data. The truncated data set represents just one statistical
sample. The
$\chi^2$ evaluation has originally nothing to do with finding a
optimal ``model/theory function'' by fitting. The minimal sum of the
weighted squares of deviations of the data from a model function
should be distinguished from $\chi^2$ and we call it $M^2$.  If one
knows the model function with certainty - this includes the knowledge
of the parameters - there is no fitting and $\chi^2 = M^2$
representing a test of statistical pureness of the data (Pearson
test). In Physics, however, one has neither a certain model/theory
function nor the certainty that the sample is statistically pure. One
has therefore to deal with a mixed and dirty situation. It may very
well be that the fits of eq.\ (\ref{eq:exp}) with $\langle r^n
\rangle$ as free fit parameters are giving a small $M^2$, but this
cannot be interpreted as a value of the $\chi^2$ distribution and
consequently as measure of the significance of the fit. The sample is
not ``true'' but just one of the possible statistical
fluctuations. Since $M^2$ is not a $\chi^2$ an estimate of errors from
such equating is an approximation. The physics constraints discussed
above have to be realized even if the $M^2$ gets worse. It just means
that the sample is not following so closely the model expectations
that the $M^2$ is absolutely minimal. Therefore it is really light
hearted to neglect the information contained in 80\% of the data and
believe that this is a valid approach. A similar remark holds for the
Fisher-Snedecor test variable in the F-test. It is recommended to
study the landmark book of Frederick James \cite{James:2006st} which was
exactly meant as an educational means for CERN users in 1970 to get
out off the over simplified application of statistics in physics. It
served as the basis for the chapters about statistics in the Review of
Particle Physics of the Particle Data Group \cite{Agashe:2014kda}
which serve as the standard in particle physics.

In the second part of the papers of Griffioen et al.\
\cite{Griffioen:2015hta} and Higinbotham et al.\
\cite{Higinbotham:2015rja} the ``continued fraction expansion'' for
the form factor $G(Q^2)$ is tried as an alternative to the many
ans\"atze in refs.\ 
\cite{Arrington:2007ux,Arrington:2015ria,Bernauer:2011zz,Bernauer:2013tpr}
yielding a radius in accord with the muonic hydrogen measurement
albeit with relatively large error. The statistical evaluation is
based on a markedly worse normalized $\chi^2/dof$. (We continue to
call it $\chi^2$ since people are so used to it.) It has to be noted
that the ``continued fraction expansion'' of $G(Q^2)$ is the only one
giving a small rms radius and was excluded from the analysis of
refs.\ \cite{Bernauer:2011zz,Bernauer:2013tpr} since it was to stiff
to fit the data. It is not better justified by any theoretical
argument than the others used, and, therefore, the ``model error''
assigned to the rms radius had to include all models with a
sufficiently good $\chi^2$ disfavoring the small radius. A detailed
discussion of the correct statistical analysis of fits to the new
Mainz cross section data (including the constant fraction expansion)
will be part of a forthcoming paper \cite{Bernauer:2015def}.

In summary, a low order expansion of the form factors has to be
consistent with our knowledge of the shape at large $Q^2$ derived from
experiments over the 50 years since the work of Robert Hofstadter. A
fit to a truncated $Q^2$-range data set cannot be used to extract a
robust value for the radius since it neglects this knowledge. The full
$Q^2$ range of $G(Q^2)$ has to be used to be able to determine the
mandatory knowledge of $\rho(r)$. Since the two papers are neglecting
this requirement they do not explain the ``proton radius puzzle''.

It is worth noting that the realization of the importance of the full
form factor also limits the conjectures of spikes or bumps at very low
$Q^2$ where no electron scattering measurements are possible. Any
structure there must introduce significant long range contribution to
the charge distribution.

{\it Acknowledgments} This work was supported by the Collaborative
Research Center 1044 and the State of Rhineland-Palatinate.
  
\bibliography{references}
 
\end{document}